\documentclass[structabstract]{aa}
\usepackage{txfonts}
\usepackage{graphicx}
\usepackage{natbib}
\bibpunct{(}{)}{;}{a}{}{,}

\begin{document}

\title{Aqueye optical observations of the Crab Nebula pulsar}

\author{German\`a, C.,\inst{1,2}, Zampieri, L.\inst{2}, Barbieri, C.\inst{3}, Naletto, G.\inst{4,5},
\v{C}ade\v{z}, A.\inst{6}, Calvani, M.\inst{2}, Barbieri, M.\inst{3}, Capraro, I.\inst{3,4}, \\
Di Paola, A.\inst{7}, Facchinetti, C.\inst{3}, Occhipinti, T.\inst{3,4}, Possenti, A.\inst{8},
Ponikvar, D.\inst{6}, Verroi, E.\inst{3,4}, Zoccarato, P.\inst{9}}
\institute{Departamento de F\'isica, Universidade Federal de Santa Catarina, Florian\`opolis, Brazil 
\and INAF-Astronomical Observatory of Padova, Italy 
\and Department of Physics and Astronomy, University of Padova, Italy
\and Department of Information Engineering, University of Padova, Italy 
\and CNR-IFN UOS Padova LUXOR, Padova Italy
\and Faculty of Mathematics and Physics, University of Ljubljana, Slovenia 
\and INAF-Astronomical Observatory of Rome, Italy
\and INAF-Astronomical Observatory of Cagliari, Italy
\and Interdipartimental Center of Studies and Activities for Space (CISAS) ``G. Colombo'', University of Padova, Italy
}


\date{Received / Accepted}

\abstract
{We observed the Crab pulsar in October 2008 at the Copernico Telescope in Asiago -- Cima Ekar with 
the optical photon counter Aqueye (the Asiago Quantum Eye) which has the best temporal
resolution and accuracy ever achieved in the optical domain (hundreds of picoseconds).}
{Our goal was to perform a detailed analysis of the optical period and phase drift of the main peak
of the Crab pulsar and compare it with the Jodrell Bank ephemerides.}
{We determined the position of the main peak using the steepest zero of the 
cross-correlation function between the pulsar signal and an accurate optical template.}
{The pulsar rotational period and period derivative have been measured with great accuracy using 
observations covering only a 2 day time interval. 
The error on the period is 1.7 ps, limited only by the statistical uncertainty.
Both the rotational frequency and its first derivative are in agreement with those 
from the Jodrell Bank radio ephemerides archive. We also found evidence of the optical 
peak leading the radio one by $\sim 230\ \mu$s. The distribution of phase-residuals 
of the whole dataset is slightly wider than that of a synthetic signal generated
as a sequence of pulses distributed in time with the probability proportional
to the pulse shape, such as the average count rate and background level are
those of the Crab pulsar observed with Aqueye.}
{The counting statistics and quality of the data allowed us to determine the pulsar period 
and period derivative with great accuracy in 2 days only.
The time of arrival of the optical peak of the Crab pulsar leads the radio one 
in agreement with what recently reported in the literature. 
The distribution of the phase residuals can be approximated with a Gaussian and is
consistent with being completely caused by photon noise (for the best data sets).}


\titlerunning{Crab pulsar optical observations with Aqueye}
\authorrunning{German\`a et al.}

\maketitle

\section{Introduction}

Since its discovery \citep{1968Sci...162.1481S,1969Natur.221..453C}, the pulsar in the Crab nebula 
has been one of the most targeted objects in the sky at all wavelengths, from radio to very high energy-rays,
serving as a test bed for pulsar theories as well as for studying astrophysical non-thermal processes.
Optical pulsations were discovered more than 40 years ago (\citealt{1969Natur.221..525C,1969ApJ...155L.121L})
and the Crab pulsar was indeed the first celestial object to be detected as a pulsating source in the optical band. 

The optical light curve of the Crab pulsar has been monitored through the years using a
variety of telescopes and instruments (e.g. \citealt{2009MNRAS.397..103S}).
The pulse shape is characterized by a double peak profile, separated in phase by
$\sim 140^{\circ}$. The shape is similar through the entire electromagnetic spectrum, although
the morphological details differ substantially from radio to gamma-rays.
Wavelength-dependent changes in the pulsar properties have been 
reported also by Percival et al. (1993) (peaks width and separation larger in the V band
than in the UV) and by \cite{2002ApJ...581..485F}.
The pulse shape is very stable (e.g. \citealt{2011AdSpR..47..365Z}), despite the secular decrease of the 
luminosity \citep{1996A&A...314..849N} and the presence of glitches and timing noise. 
Occasionally small variations of the shape of the pulse have been observed 
\citep{2007Ap&SS.308..595K}.

Several issues concerning the pulsar engine and the actual geometry of the emission regions are still debated,
ranging from the nature and location of the acceleration mechanism, to wavelength 
dependent variations of the pulse profile, to properties of the Giant Radio Pulses 
(GRPs). In particular, the study of GRPs is currently a very active field, with deep 
theoretical implications. So far GRPs have been observed in a handful of pulsars. 
However, an analogous optical phenomenon has been observed only at the Crab. 
GRPs seem to show a weak correlation with optical pulses 
(\citealt{2003A&A...411L..31K,2004ApJ...605L.129R}), which are on average 3\% 
brighter when coincident with GRPs (\citealt{2003Sci...301..493S}).
Recent coeval timing at optical and radio wavelengths by \cite{2008A&A...488..271O}
found a 255 $\pm$ 21 $\mu$s delay of radio with respect to optical pulse.



In the last few years we started a monitoring programme of the
Crab pulsar in the optical band aimed at studying the long term stability and sub-$\mu$s structure
of its pulse shape, and at performing accurate optical timing of the
main peak. By comparing the behaviour in other wavebands, especially radio,
we aim to improve the understanding of the geometry of the acceleration site.
Observations are performed by means of a very fast single photon-counter 
instrument, Aqueye, mounted at the 182 cm Copernico Telescope in Asiago \citep{2009JMOp...56..261B,2011AdSpR..47..365Z}.
The design of Aqueye follows that of QuantEYE \citep[the Quantum EYE;][]{2005astro.ph.11027D,2008JMO..11..190B}, 
an instrument specifically tailored for studying rapid optical variability of astrophysical 
sources with the ESO E-ELT. A second version of the instrument, named Iqueye, has been 
installed and successfully used at the ESO NTT telescope \citep{2009A&A...508..531N,2010SPIE.7735E.138N}.

In a preliminary investigation \citep{2011AdSpR..47..365Z} we concentrated on the pulse 
shape stability of the Crab pulsar and found  
that it is stable at the level of $\sim 1$\% on a timescale of 14 years. This result 
reinforces evidence for decadal stability of the inclination angle between the spin 
and magnetic axis, and of the thickness of the emission region.
Here we present a follow-up investigation reporting accurate phase analysis 
of optical timing of the main peak and comparing it with radio ephemerides 
of the Jodrell Bank (JB) radio Observatory \citep{1993MNRAS.265.1003L}.

The plan of the paper is as follows. In section~\ref{observations} we list observations 
of the Crab pulsar performed with the Aqueye instrument and discuss the barycenterization procedure. 
In Section~\ref{phaseanalysis} we illustrate the implementation of phase analysis. 
In Section~\ref{results} our results are presented and compared with radio ephemerides. 
Section~\ref{conclusions} summarizes the conclusions.

\begin{table}
      \caption[]{Log of October 2008 Crab pulsar observations performed 
      with Aqueye mounted at the 182cm Copernico telescope in Asiago. The start time 
      of the observations is the GPS integer second, accurate to $\pm$30 nanoseconds.
      }
         \label{tab1}
     $$ 
         \begin{array}{p{0.05\linewidth}p{0.35\linewidth}p{0.1\linewidth}}
                     \hline
            \noalign{\smallskip}
            & Starting time &  Duration\\
            & (UTC)         &  (s)     \\
            \noalign{\smallskip}
            \hline
            \noalign{\smallskip}
            1 & October 10, 23:45:14  &  898   \\
            2 & October 11, 00:05:07  &  1197  \\
            3 & October 11, 01:00:22  &  1797  \\            
            4 & October 11, 01:45:44  &  1797  \\
            5 & October 11, 02:23:07  &  1631  \\
            6 & October 11, 03:23:46  &  1197  \\
            7 & October 11, 23:08:03  &  292   \\                                                          
            8 & October 11, 23:25:09  &  3597  \\
            9 & October 12, 00:54:31  &  1794  \\
           10 & October 12, 23:03:59  &  292   \\
           11 & October 12, 23:13:57  &  3998  \\ 
           12 & October 13, 00:57:07  &  7194  \\
            \noalign{\smallskip}
            \hline
         \end{array}
     $$ 
\end{table}

\section{Observations}
\label{observations}

The Crab pulsar was observed with Aqueye mounted at the 182 cm Copernico Telescope in Asiago.
The observations were performed in 2008 and lasted for three nights, starting from October 10. 
The sky was clear and seeing conditions fair (1.5 arcsec average). For a timing log of observations, see Table~\ref{tab1}. 
During each observing run we recorded the arrival time of $\sim 0.15-3.6 \times 10^7$ photons, 
time-tagged with a relative time accuracy of $\sim$100 picoseconds and an
absolute precision (referred to UTC) better than 500 ps (for details about 
the timing accuracy of the acquisition system see \citealt{2009A&A...508..531N}).
To our knowledge, this is the most accurate measurement of photon arrival times
from the Crab pulsar ever obtained in the optical band.

The time-tag of each detected photon in the unbinned time series was reduced to the Solar System
barycentric time using the software Tempo2\footnote{http://www.atnf.csiro.au/research/pulsar/ppta/tempo2}
\citep{2006MNRAS.369..655H,2006MNRAS.372.1549E}.
The adopted position of the Crab pulsar is that reported in the Jodrell
Bank monthly ephemerides (R.A. 05h 34m 31.97232s, DEC. +22$^0$ 00$'$ 52.0690$''$ [J2000]), with no correction for proper motion.
To perform this conversion the software needs also accurate value of the observatory
geocentric coordinates. 
They were obtained with a GPS receiver, which was connected to an antenna 
with a length compensated cable and situated at the dome of the telescope. 
At least 6 GPS satellites were used in the positional data acquisition, which 
typically lasted 3 hours and was repeated for several days. Finally the position 
of the antenna was referred to the intersection of the telescope hour angle and 
declination axes by laser assisted metrology. We estimate the positional error 
to be $\sim$30 cm, amply sufficient for the purpose of this paper.

In order to compare our ephemerides to those reported in the JB Observatory 
radio archive we baricentered the time-tags also in the Tempo1 emulation mode.
In doing so, we found an error in the actual value of the Roemer delay, caused by
some inconsinstency in the Earth configuration files related to the calculation
of the polar motion of the Earth\footnote{The Roemer delay computed in Tempo2 showed an anomalous
oscillation at around the time of our observations,
reaching a maximum value of 30 $\mu$s (instead of $\la 35$ ns; see e.g. \citealt{2006MNRAS.369..655H}).
This caused a drift of the phase of the main peak of $\sim$1 ms per day and a lengthening of the rotational 
period of $\sim$0.4 ns.}.
This problem was solved by using updated Earth configuration files loaded from the Tempo2 SCM 
repository\footnote{http://tempo2.cvs.sourceforge.net/viewvc/tempo2/tempo2/T2runtime/}.

\begin{table}
      \caption[]{Geocentric coordinates of the 182 cm Copernico telescope in Asiago.
      The $3\sigma$ uncertainty is 0.3 m.}
         \label{tabcoords}
     $$ 
         \begin{array}{p{0.3\linewidth}p{0.3\linewidth}p{0.3\linewidth}}
                     \hline
            \noalign{\smallskip}
            $x$ & $y$ & $z$ \\
            (m) & (m) & (m) \\
            \noalign{\smallskip}
            \hline
            \noalign{\smallskip}             
            4360966.0  &  892728.1  &  4554543.1  \\
            \noalign{\smallskip}
            \hline
         \end{array}
     $$ 
\end{table}

\section{Phase-analysis}
\label{phaseanalysis}

We analyzed the evolution of the phase of the main peak of the Crab pulsar using
the barycentered event list.
A reference period $P_{init}$ is assumed and the light curve is divided into
$n$ seconds long segments. Each segment is then folded over $P_{init}$ and
is binned at $\sim$1/300 in phase. The phase of the main peak is determined by 
cross-correlating the pulse shape with a template, as summarized in the Appendix. 
Our method is conceptually similar to the one adopted by \citet{2008A&A...488..271O} 
and is more accurate than previous approaches based on fitting the main peak 
with a simple analytic function like a Lorentzian, Gaussian or parabola 
(see e.g. \citealt{2006A&A...456..283O}), given its asymmetric shape.


Following the standard pulsar spin down model, we describe the phase drift 
of the main peak with respect to uniform rotation using a third-order polynomial,
i.e.:
\begin{equation}
\Delta\phi(t)=\phi(t)-\phi'(t)=\phi_{0}+(\nu-\nu_{init})(t-t_{0})+\frac{1}{2}\dot{\nu}(t-t_{0})^{2}+\frac{1}{6}\ddot{\nu}(t-t_{0})^{3}
\label{eq1}
\end{equation}
where $t_0$ is a reference time,
$\phi_0=\phi(t_0)$ is the phase of the main peak at $t_0$, $\phi'(t)=(t-t_{0})/P_{init}$ is the phase 
for constant rotation at frequency $\nu_{init}=1/P_{init}$, and $\nu$, $\dot{\nu}$, 
$\ddot{\nu}$ are the actual rotational frequency and its first and second derivatives, respectively. 
$P_{init}$ is chosen in such a way that $\Delta \phi$ varies slowly during the period of observation.
For a 2 days baseline, the linear and quadratic terms, i.e. the first and second derivatives of the phase, are sufficient to describe the drift, and we
can safely neglect the cubic term in equation~(\ref{eq1}).
Then, the expression for $\Delta \phi(t)$ (eq.~[\ref{eq1}]) becomes of the form
\begin{equation}
\psi(t)=\phi_{0}+a(t-t_{0})+b(t-t_{0})^{2} \, ,
\label{eq2}
\end{equation}   
where $\phi_0$, $a$ (in units of $s^{-1}$) and $b$ (in units of $s^{-2}$) 
are determined by fitting $\psi(t)$ to the observed phase-drift.
After determining $\psi(t)$ from the fit, the phase of the main peak is given by:
\begin{equation}
\label{eq4}
\phi(t)=\phi'(t)+\Delta\phi(t)=\nu_{init}(t-t_{0})+\psi(t) \, .
\end{equation}

\subsection{The radio phase from the Jodrell Bank ephemerides archive}

We compared the phase of the Crab pulsar measured by Aqueye with that reported in the 
radio archive at the JB Observatory\footnote{http://www.jb.man.ac.uk/$\sim$pulsar/crab.html} 
\citep{1993MNRAS.265.1003L}. The phases of the JB ephemerides are those 
of the main peak at infinite frequency at the barycenter of the Solar System.
The observed barycentric radio phase $\phi_r(t)$ is obtained using the values of $\nu_r$, $\dot{\nu}_r$ and $\ddot{\nu}_r$ 
nearest to our observing epochs and reported in the archive\footnote{http://www.jb.man.ac.uk/$\sim$pulsar/crab/all.gro}.
The radio phase drift is given by an expression similar to equation~(\ref{eq1}):
\begin{equation}
\Delta\phi_r=\phi_r(t)-\phi'(t)=\phi_{r,0}+(\nu_r-\nu_{init})(t-t_{0})+\frac{1}{2}\dot{\nu}_r(t-t_{0})^{2}
+\frac{1}{6}\ddot{\nu}_r(t-t_{0})^{3}
\label{eq6}
\end{equation}
The optical phase is in agreement with the radio one if the phase drifts inferred 
from equations~(\ref{eq1}) and~(\ref{eq6}) are in agreement. 
The radio phase $\phi_{r,0}$ at epoch $t_{0}$ is calculated by means of a quadratic 
extrapolation starting from the closest value in time reported in the JB radio ephemerides 
(Oct 15, 2008) and is in agreement, within the errors, with the value calculated with a Fortran 
code available at the JB radio ephemerides website.
No known glitch (down to the intensity to which radio monitoring is sensitive to) affected 
the Crab in the interval of time between the determination of the
Crab radio parameters and the optical observations.


\section{Results}
\label{results}

\begin{figure}
\resizebox{\hsize}{!}{\includegraphics[angle=0]{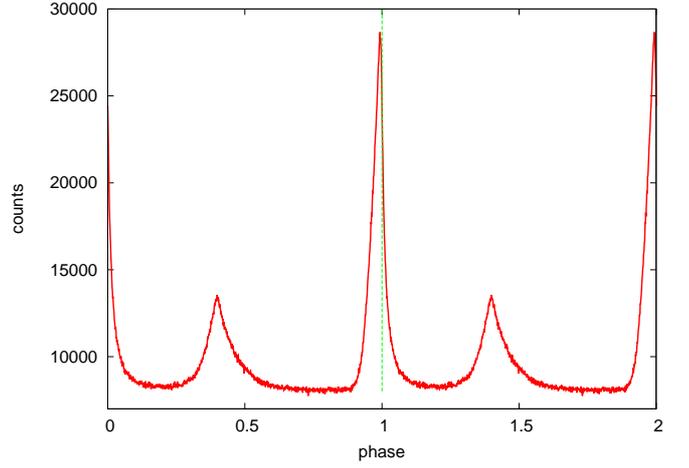}}
\caption{Folded light curve of the Crab pulsar as a function of phase for the Aqueye 
observation 4 in Table~\ref{tab1}. The folding period and the bin time are 0.0336216417 s
and $33.6 \, \mu$s, respectively.
The typical double peak profile of the pulse is recognizable. For sake of clarity two 
rotations of the neutron star are shown. Phase zero/one corresponds to the position
of the main peak in the radio band and is marked with a vertical {\it dashed} line.}
\label{fig3}
\end{figure}

In Figure~\ref{fig3} we show the light curve of the Crab pulsar folded over the 
average spin period for one of the Aqueye observations (see also \citealt{2011AdSpR..47..365Z}). 
The light curve includes the contribution of both the pulsar and the nebular background
entering the Aqueye pinhole entrance aperture. For a whole observation the counting statistics is large 
and hence the bin time is smaller than that adopted for the 
phase analysis, for which the typical integration time is a few seconds (see below).
The average count rate of the Crab pulsar (all channels) measured by Aqueye is $\sim 5500$ 
counts s$^{-1}$. The count rate of the background, estimated from the off-pulse region of the
folded light curve (see Figure~\ref{fig3}), is $\sim 4500$ counts s$^{-1}$, 
which implies a total number of net source photons of $\sim 2.6\times 10^6$,
time tagged to better than 500 ps with respect to UTC. As we already point out in
\citet{2011AdSpR..47..365Z}, the pulse shape agrees well with the 33 years-old 
pulse profile obtained by \citet{1975ApJ...200..278G}, as well as with the more 
recent one by \citet{2007Ap&SS.308..595K}.

\begin{figure}[!ht!]
\begin{center}
\resizebox{\hsize}{!}{\includegraphics[width=1\textwidth]{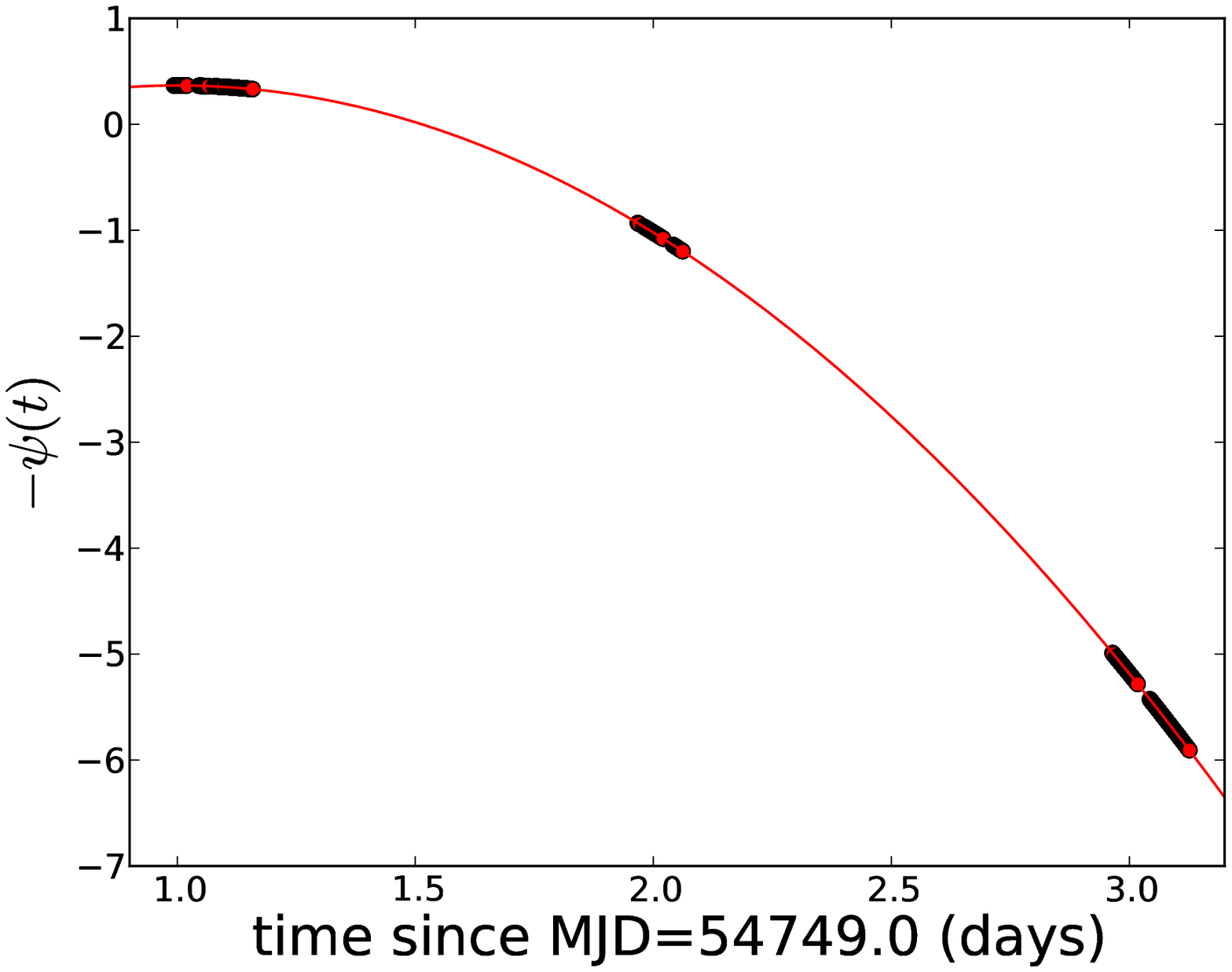}}
\resizebox{\hsize}{!}{\includegraphics[width=1\textwidth]{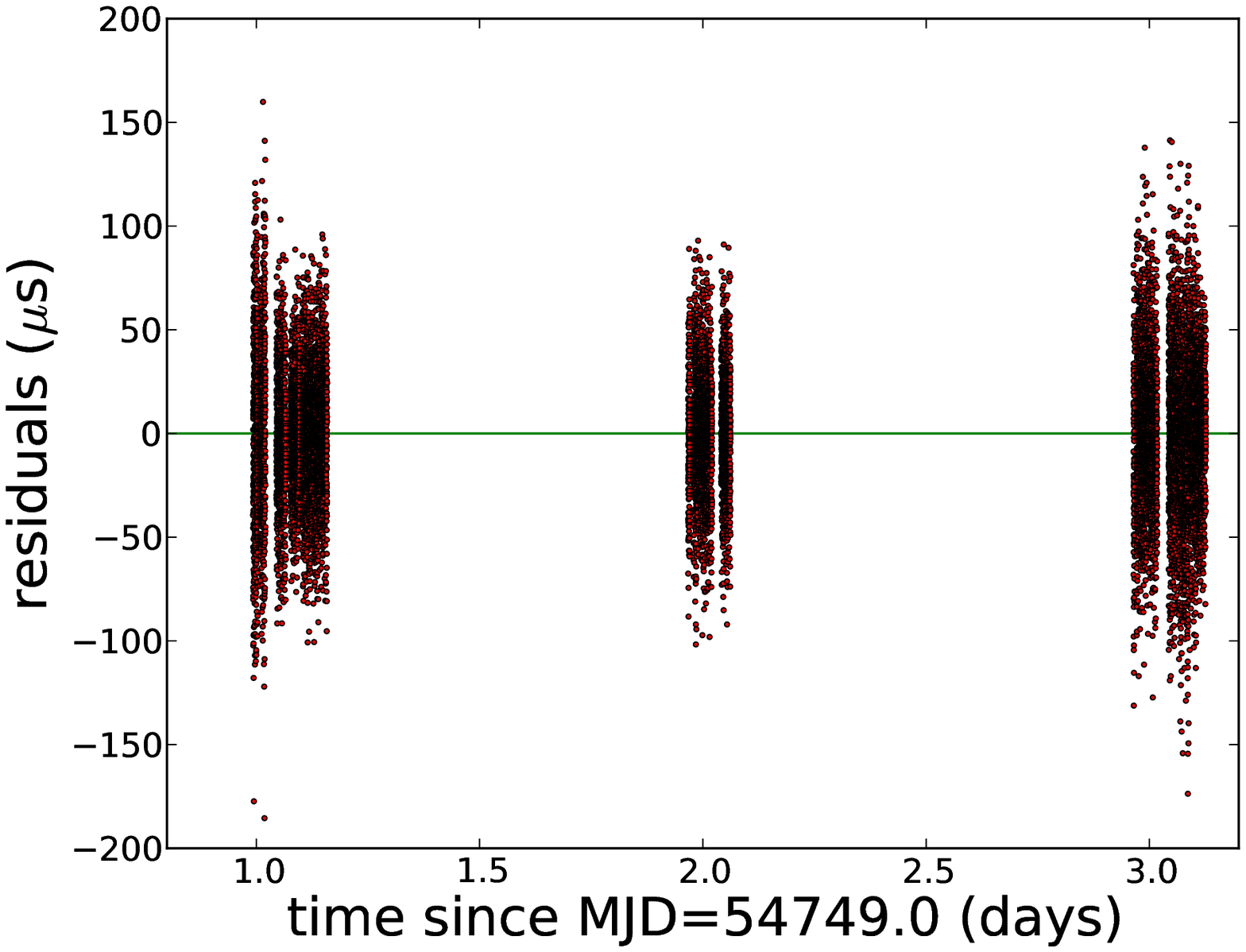}}
\caption{\emph{Top panel}: Phase-drift of the main peak of the Crab pulsar 
(changed sign) measured during the observing run in Asiago in October 2008. 
The (red) curve is the best-fitting parabola (eq[\ref{eq2}]).
The reference epoch $t_0$ is MJD=54749.0, while the reference rotational 
period is $P_{init}=0.0336216386529$ s.
\emph{Bottom panel}: Phase residuals (in $\mu$s) after subtracting the 
best-fitting parabola to the phase-drift.}
\label{fig6}
\end{center}
\end{figure}

\subsection{Optical phase drift}
\label{optphaseanalysis}

In the following we focus on the detailed analysis of the optical phase drift of the main peak
of the Crab pulsar and compare it with the behaviour observed in the radio using the JB ephemerides.
Figure~\ref{fig6} shows the best-fitting parabola to the phase-drift of the Crab pulsar
measured by Aqueye. The phase of the main peak is calculated using intervals 2 seconds long.
The typical 1-$\sigma$ uncertainty on the position of the peak is $\sim 30 \, \mu$s.
The best-fit gives a reduced $\chi^2 \sim 1.06$. The reference epoch $t_0$ is MJD=54749.0, while the reference rotational 
period $P_{init}=1/\nu_{init}$ used to fold the light curve is $P_{init}=0.0336216386529$ s.
The best-fitting parabola of the optical phase drift $\psi(t)$ (eq.~[\ref{eq2}]) is:
\begin{eqnarray}
\psi(t) &=& (1.021431 \pm  0.000081) \nonumber \\
&-& [(3.21329 \pm 0.00011)\times 10^{-5} \, s^{-1}] \, (t-t_0) \nonumber \\
&+& [(1.859380 \pm 0.000029) \times10^{-10} \, s^{-2}] \, (t-t_0)^{2} \, .
\label{eq7}
\end{eqnarray}
The quoted uncertainties are the 1-$\sigma$ errors for one interesting parameter.
Figure~\ref{fig6} shows also the phase residuals remaining after subtracting equation~(\ref{eq7}) 
from the measured phase drift. The distribution shows a spread of $\la 100 \, \mu$s
($\sim$ 0.003 cycles) and is rather symmetric around zero (see Figure~\ref{fig8}), 
testifying that all the observations are phase-connected.

The rotational frequecy is $\nu=d\phi(t)/dt$. Thus,
inserting equation~(\ref{eq7}) into equation~(\ref{eq4}) and taking the 
derivative with respect to $t$, we obtain an expression for the rotational
frequency $\nu$ and period $P=1/\nu$ at a given (barycentered time) $t$. Table~\ref{tab2} 
lists the rotational periods of the Crab pulsar measured with Aqueye and 
those from the JB ephemerides at 3 reference dates (barycentered
MJD=54750, 54751, 54752). The differences between optical and radio ranges
from $\sim 2$ to 4 ps. The statistical error on the optical rotational 
periods is 1.7 ps ($1\sigma$ error).
The quoted error from the JB radio archive is $\sim 0.1$ ps. Thus, 
the rotational periods calculated by Aqueye agree within the statistical 
error with those derived from radio measurements. Also the measurements of
the first derivative of the rotational frequency ${\dot \nu}$ are in 
agreement within the errors: ${\dot \nu}_{Aqueye} = 3.71876\times10^{-10} \pm
6\times10^{-15}$ s$^{-2}$ and ${\dot \nu}_{JB} = 3.718655\times10^{-10}
\pm 2\times10^{-16}$ s$^{-2}$.

We emphasize that, in order to compare our data with those of the JB 
radio ephemerides, we barycentered the time-tags of the optical photons
in Tempo2 using the Tempo/Tempo1 emulation mode. This is because, for
historical reasons, the JB radio ephemerides are calculated using Tempo, 
which is the older version of the software used for barycentering. 
The systems of time adopted in the two packages are different.
Tempo2 uses the Barycentric Coordinate Time (TCB), while Tempo the 
barycentric dynamical time (TDB)\footnote{TCB is a coordinate time referred to the barycenter of the Solar System,
synchronized with the proper time of a distant observer comoving with it. 
The system adopted in Tempo1 is the barycentric dynamical time (TDB), effectively measured 
in units that differ subtly from the conventional SI second \citep{2006MNRAS.369..655H}.
It is as if the time dilation effects
were not correctly accounted for using TDB units, so that, for example,
rotational periods in the TDB system are systematically shorter than the TCB ones.}. 
In Table~\ref{tab3} we report rotational periods after barycentering 
with Tempo2 in TCB units. Although there are other differences between the
Tempo/Tempo1 mode setup used for calculating the JB ephemerides of the Crab and 
the full Tempo2 mode (as for example the adopted Solar System ephemerides),
the main difference between rotational periods reported in Tables~\ref{tab2}
and~\ref{tab3} is due to the use of TCB units (SI units) instead of TDB.
The rotational periods of the Crab are $\sim 0.5$ ns longer than those 
measured using Tempo1 in TDB units. If we take the ratio of the periods 
in Table~\ref{tab3} (TCB units) to those in the second column of Table~\ref{tab2} 
(TDB units) we find that the ratio of the two time units is
$K \sim 1 + 1.53 \times 10^{-8} \pm 1.3 \times 10^{-10}$,
consistent with the value reported by \cite{1999A&A...348..642I} and 
\cite{2006MNRAS.369..655H}. The constant $K$ sums up a contribution 
from the linear term of the Einstein delay, $L_C$, and another term 
from the gravitational plus spin potential of the Earth, $L_G$.
Thanks to its timing capability and performances,
Aqueye can put in evidence the occurrence of the corrections $L_C$ and $L_G$ 
to the pulsar spin period in only two days of data taking on the Crab pulsar.

\begin{table}[!t!]
      \caption{\footnotesize{Rotational periods of the Crab pulsar measured 
      by Aqueye in 2008 compared to those reported in the Jodrell Bank radio 
      ephemerides. The time-tags were barycentered 
      in Tempo1 emulation mode.}}
         \label{tab2}
     $$ 
         \begin{array}{p{0.35\linewidth}p{0.3\linewidth}p{0.3\linewidth}}
                     \hline
            \noalign{\smallskip}
            MJD$^a$  &  $P$(Aqueye)$^b$     & $P$ (JB) \\
                     &  (s)                 & (s)                  \\
            \noalign{\smallskip}
            \hline
            \noalign{\smallskip}
            54750.0  &  0.033621638649  &  0.033621638653  \\
            54751.0  &  0.033621674970  &  0.033621674973  \\
            54752.0  &  0.033621711290  &  0.033621711292  \\
            \noalign{\smallskip}
            \hline
         \end{array}
     $$ 
\begin{list}{}{}
\item[$^{\mathrm{a}}$] MJD at the solar system barycenter (Tempo1 mode).
\item[$^{\mathrm{b}}$] $\sigma_P$=1.7 ps (68\% statistical error)
\end{list}
\end{table}

\begin{table}[!ht!]
      \caption{\footnotesize{Rotational periods of the Crab pulsar measured by Aqueye 
      in October 2008. The time-tags were barycentered in Tempo2 (TCB units).}}
         \label{tab3}
     $$ 
         \begin{array}{p{0.35\linewidth}p{0.3\linewidth}}
                     \hline
            \noalign{\smallskip}
            MJD$^a$  &  $P$ (Aqueye)$^b$ \\
                     &  (s)   \\
            \noalign{\smallskip}
            \hline
            \noalign{\smallskip}
            54750.0  &  0.033621639166 \\
            54751.0  &  0.033621675484  \\
            54752.0  &  0.033621711803  \\            
            \noalign{\smallskip}
            \hline
         \end{array}
     $$ 
\begin{list}{}{}
\item[$^{\mathrm{a}}$] MJD at the solar system barycenter (Tempo2 mode).
\item[$^{\mathrm{b}}$] $\sigma_P$=1.7 ps (68\% statistical error).
\end{list}
\end{table}

\begin{figure}[!t!]
\begin{center}
\resizebox{\hsize}{!}{\includegraphics[width=0.8\textwidth]{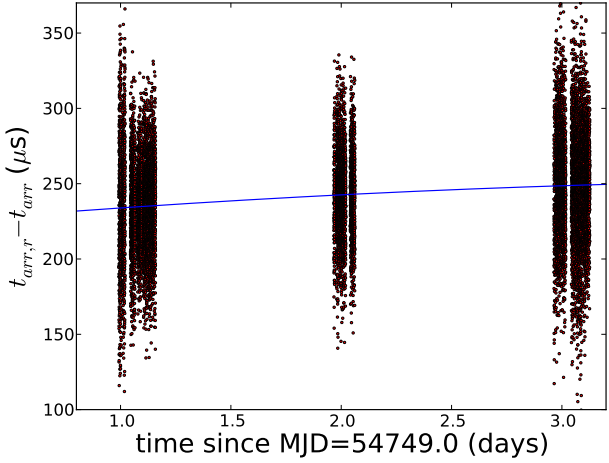}}
\caption{Difference between the optical and radio time of arrival of the main 
peak of the Crab pulsar. The optical
peak leads the radio one by $\sim 230\ \mu s$ (at MJD=54750, epoch of the first observation). 
The (blue) line is the radio-optical drift, which is consistent with zero within the errors (see text).
}
\label{fig7}
\end{center}
\end{figure}

\begin{figure}[!t!]
\begin{center}
\resizebox{\hsize}{!}{\includegraphics[width=0.8\textwidth]{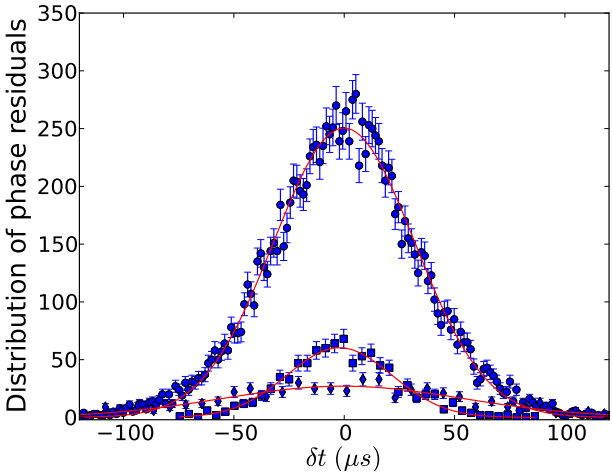}}
\caption{Distribution of the residual of the optical phase of the
main pulse around the best fit shown in Figure~\ref{fig6}. 
Each point represents the number of phase residuals $N_{i}$ in the $i$-th bin. 
The error bar on each bin is $\sqrt{N_{i}}$.
The three datasets refer to all the observations ({\it circles}), to observation
4 ({\it squares}) and observation 1 ({\it diamonds}), respectively. 
The bin widths are $1.5\mu$s, $4.1\mu$s and $7.3\mu$s.
The (red) solid lines superimposed to each dataset represent the best fitting 
gaussian with $\sigma \sim 32 \, \mu$s ({\it circles}), $\sigma \sim 24 \, \mu$s 
({\it squares}) and $\sigma \sim 54 \, \mu$s ({\it diamonds}), respectively.}
\label{fig8}
\end{center}
\end{figure}

\begin{figure}[!t!]
\begin{center}
\resizebox{\hsize}{!}{\includegraphics[width=0.8\textwidth]{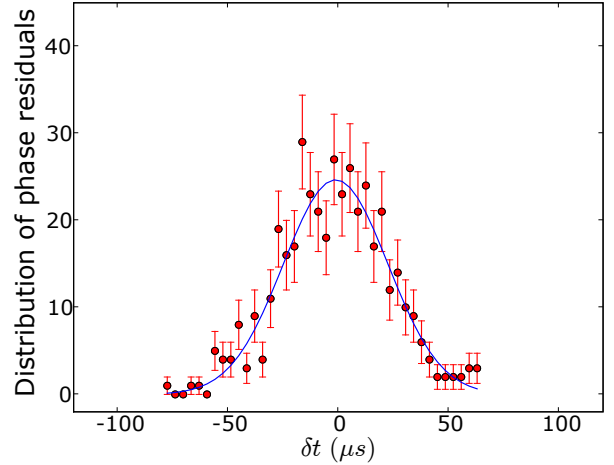}}
\caption{Same as Figure~\ref{fig8} for a simulated signal 
with superimposed random noise (see text for details). The assumed count rate is the
average count-rate measured by Aqueye. The bin width is $3.6\mu$s. The distribution 
is fit by a gaussian with $\sigma \sim 24 \, \mu$s.}
\label{fig9}.
\end{center}
\end{figure}

\subsection{Radio-Optical delay}

The time of arrival at the detector of the first pulse of the light curve after a
certain epoch $t$ is $t_{arr}=\phi P_{init}$,
where $\phi$ is the phase defined in Section~\ref{phaseanalysis}. Figure~\ref{fig7} shows the difference 
between the optical time of arrival of the main peak of the Crab pulsar and the radio
one determined from the JB radio ephemerides. The dispersion measure at around the
epoch of the Aqueye observations was 56.7842 pc cm$^{-3}$ (Oct 15, 2008).
We find that the optical peak leads the radio one. Taking into account the uncertainty on the time of
arrival quoted in the JB archive\footnote{http://www.jb.man.ac.uk/~pulsar/crab/crab2.txt} ($\sim60\ \mu$s)
and the errors from the fit, the time difference is $\sim 230 \pm 60 \, \mu s$ at MJD=54750,
with a drift of $\simeq 7 \, \mu s$/day. While within the errors the drift is consistent 
with zero, the difference in the arrival times is significant. The optical peak leads
the radio one, in agreement with what was found previously by \citet{sanwal99},
\citet{2006A&A...456..283O} and \citet{2008A&A...488..271O}, the latter obtained 
using simultaneous optical and radio observations. Our value 
of the radio delay is also consistent with the recent measurement performed
by our group with Iqueye (178 $\mu$s; \citealt{2012IAUS..285..296C}), but larger
than the one reported in \citet{2003Sci...301..493S}. 
The uncertainity on our measurement is dominated by the error on
the radio ephmerides, and can be further reduced in future using simultaneous 
radio-optical observations.

\subsection{Phase noise}
\label{noise}

Figure~\ref{fig8} shows the distribution of the phase residuals in Figure~\ref{fig6}.
The distribution is fit with a gaussian and, for all the observations, 
gives a reduced $\chi^2 \sim 0.9$.
The gaussian has $\sigma \sim 32 \, \mu$s, consistent with the error bar of each measured phase
(see Section~\ref{phaseanalysis}). This indicates that, with
the present accuracy, the phase noise of the Crab pulsar observed with Aqueye 
can be approximated with a Gaussian. However, the distribution of phase residuals in different observations 
appears to have different widths. The distributions with the smallest (obs. 4) and largest
(obs. 1) widths are also shown in Figure~\ref{fig8} for comparison. This suggests the existence
of additional errors in the data chain of some observations, possibly induced
by signal loss (e. g. clouds or telescope tracking errors).


We compared the observed phase residuals with those obtained from a synthetic signal
generated as a sequence of pulses distributed in time with the probability proportional
to the pulse shape, such as the average count rate and background level are
those of the Crab pulsar observed with Aqueye.
The signal has superimposed random noise and lasts $\sim 850$ s. 
Figure~\ref{fig9} shows the distribution of the residuals of the simulated signal. 
The distribution is clearly Gaussian but with $\sigma\sim 24 \, \mu$s. This
is smaller than the total phase noise distribution in Figure~\ref{fig8}.
However, as mentioned above, the observations have different widths of the
distributions. Some of them are comparable (obs. 4) or marginally larger (obs. 6 and
7) than that inferred from the simulation, while some others have more outliers.
It is thus possible that the outliers are due to errors in the data chain, induced 
by the smaller quality of the dataset. For the best dataset (obs. 4; see Figure~\ref{fig8}),
the measured width of the phase residuals distribution appears to approach
the theoretical expectations for phase noise induced by pure photon statistics.

\section{Discussion and conclusions}
\label{conclusions}

We observed the Crab pulsar with the photon counting instrument Aqueye, 
mounted at the 182cm Copernico telescope in Asiago, during the nights of October 10-13, 2008.
The counting statistics and quality of the data allowed us to monitor the 
phase of the main peak of the Crab pulsar over 2~s long intervals and to determine 
the pulsar rotational period and period derivative with great accuracy, using 
observations covering only a 2 day interval in time. The statistical error on the
period inferred from a fit of the pulsar phase drift is of the order of a few picoseconds. 
The measurements of the period and period derivative agree within the statistical 
error with those inferred from the JB ephemerides.


We also found that the time of arrival of the optical peak of the Crab pulsar
leads the radio one (with the latter inferred from the JB radio ephemerides)
in agreement with previous findings \citep{sanwal99,2003Sci...301..493S}. 
The actual value of the radio delay, $\sim 230 \, \mu$s, is in agreement 
with the most accurate measurement previously reported in the literature by 
\citet{2008A&A...488..271O}. Previous measurements performed using fitting 
functions of the main peak are less precise. In such a case, as pointed by 
\citet{2006A&A...456..283O} and \citet{2012IAUS..285..296C}, the phase of the 
main peak depends on the chosen fitting function, introducing a systematic difference
caused by the intrinsic asymmetric shape of the peak.
Assuming a Gaussian or Lorentzian fit, different values of the phase of the peak
are obtained for different fitting ranges around it. Using a 0.1, 0.05, 0.02 phase 
interval to the right and to the left of the peak, the difference in position is
$247 \, \mu$s, $230 \, \mu$s, $134 \, \mu$s for the Gaussian fit
and $197 \, \mu$s, $197 \, \mu$s, $114 \, \mu$s for the Lorentzian fit,
respectively.
Furthermore, it is important to note that the correct estimate of the 
radio-optical time delay can be obtained only from simultaneous optical 
and radio observations using similar procedures for the barycenterization 
and the analysis (e.g. \citealt{2012IAUS..285..296C}).

As pointed out by \citet{2006A&A...456..283O}, a time delay of the radio peak 
of $\sim 230 \, \mu$s could have two different interpretations. The emission
region of the optical radiation is: (a) higher in the magnetosphere ($\sim 70$ km) than 
the radio emission, (b) located at a different angle ($\sim 2.5^0$) with respect to the radio one.

We also studied the phase noise distribution of the Crab pulsar observed with Aqueye and found
that, with the present accuracy, it can be modelled as a Gaussian.
The width of the distribution of the entire dataset is slightly larger than that induced 
by pure photon random noise from a synthetic signal having the same pulse shape, average 
count rate and background level of the Crab pulsar observed with Aqueye.
However, residual systematic errors in the data chain may be present in a subset
of observations that tend to broaden the distribution. In fact, the distribution
of the best batch of data, i.e. that with the smaller width, is consistent with 
that induced by photon statistics. While the observed broadening seems to be
caused mostly by the deterioration in the quality of some observations, the existence 
of a smaller source of phase noise, possibly related to the intrinsic pulsar mechanism,
cannot be ruled out at present and needs to be carefully investigated with
future observations.


\section{Acknowledgments}

We thank the referee for his/her constructive criticisms that helped to 
improve our paper.
We would like to thank also
Alessandro Patruno (Astronomical Institute, University of Amsterdam) for useful discussions.
We acknowledge the use of the Crab pulsar radio ephemerides available at the web site of the
Jodrell Bank radio Observatory (http://www.jb.man.ac.uk/$\sim$pulsar/crab.html; \citealt{1993MNRAS.265.1003L}).
This work has been partly supported by the University of Padova, by the Italian Ministry 
of University MIUR through the program PRIN 2006 and by the Program of Excellence 
2006 Fondazione CARIPARO.

\bibliographystyle{aa} 
\bibliography{biblio}

\appendix
\section{Calculating the phase of the main peak}

We start folding the light curve over a fixed reference period $P_{init}$, 
corresponding to the period of the pulsar some time during the complete observing run. 

\begin{figure}[h]
\includegraphics[angle=-90,width=200pt]{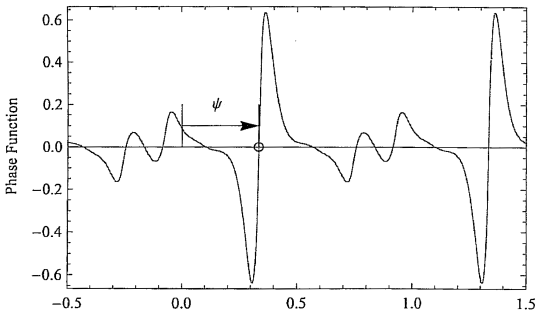}
\caption{Example of the behaviour of $K^{'}(t_1,x)$.}\label{Fig400}
\end{figure}

The phase drift of the main peak $\psi$ (see Section~\ref{phaseanalysis}) is obtained by finding the maximum of the correlation function $K$ between the actual pulsar signal $P(\phi(t))$ and a template $S(t/P_{init}-x)$ with a time dependence closely resembling  the actual pulsar signal, but with a varying phase $x$. 
The function $S(x)$ is a periodic function with period 1, obtained as a smoothed version of the whole dataset (using 3504 data points in phase):
\begin{equation}
K(t_1,x)=\int_{t_1}^{t_1+\Delta t}\mathrm{P}(\phi(t))S(t/P_{init}-x)dt
\label{app_eq1}
\end{equation}
where $\Delta t$ is the averaging time (usually 2 seconds) and $t_1+\Delta t/2$ the mid point of the averaging interval. After taking the derivative of $K$ with respect to $x$, we obtain:
\begin{equation}
K^{'}(t_1,x)= - \int_{t_1}^{t_1+\Delta t}\mathrm{P}(\phi(t))S^{'}(t/P_{init}-x)dt
\label{app_eq2}
\end{equation}
The value of $\psi$ is obtained taking $K^{'}(t_1,x)=0$ and considering only the zero with the steepest crossing (see Fig.~\ref{Fig400}).

We select the steepest zero crossing because we are looking for the maximum which corresponds to the sharpest peak of $K$. Let $f(x)$ be a differentiable function with many extrema $x_1$, $x_2$,~\dots. Then, in the vicinity of each peak it can be expanded as $f(x)=f(x_k)+(1/2)f''(x_k)(x-x_k)^2+ \dots$. The function has maximum at $x_k$ if $f''(x_k)<0$ and a minimum if $f''(x_k)>0$. $f'(x)$ has a zero at $x_k$ and the slope of $f'(x_k)$ is $f''(x_k)$. The absolute value of $f''(x_k)=f(x_k)/w^2$, where $w$ is the width of the peak of the function at half maximum. Since it is unlikely that a correlation function would have a small very narrow peak that would make $|f''|$ larger than the main peak, one concludes that the highest peak also has the largest value of $|f''|$ and hence the steepest $|f'|$.

\end{document}